\documentclass[showpacs,aps,prl,twocolumn,nofootinbib]{revtex4}
\usepackage{amsmath,axodraw4j,epsfig,color}
\usepackage[normalem]{ulem}

\newcommand{\ie}{{\it i.e.}}
\newcommand{\eg}{{\it e.g.}}
\newcommand{\delphes}{{\sc Delphes}}
\newcommand{\madgraph}{{\sc MadGraph}}
\newcommand{\madanalysis}{{\sc MadAnalysis}}
\newcommand{\feynrules}{{\sc Feyn\-Rules}}
\newcommand{\pythia}{{\sc Pythia}}
\newcommand{\be}{\begin{equation}}
\newcommand{\ee}{\end{equation}}
\def\bsp#1\esp{\begin{split}#1\end{split}}
\newcommand{\met} {\ensuremath{E\!\!\!\!/_T}}

\begin{document}

\leftline{}
\rightline{IPHC-PHENO-13-03; CERN-PH-TH/2013-078}

\title{Probing top anomalous couplings at the LHC with trilepton signatures\\in the single top mode}

\author{
  Jean-Laurent Agram$^{a}$,
  Jeremy Andrea$^{b}$,
  Eric Conte$^{a}$,
  Benjamin Fuks$^{b,c}$, 
  Denis Gel\'e$^{b}$,
  Pierre Lansonneur$^{b}$}

\affiliation{
$^{(a)}$ Groupe  de Recherche de Physique des Hautes Energies (GRPHE),
    Universit\'e de Haute-Alsace, IUT Colmar, 34 rue du Grillenbreit BP 50568, 
    68008 Colmar Cedex, France\\
$^{(b)}$ Institut Pluridisciplinaire Hubert Curien/D\'epartement Recherches Subatomiques, 
    Universit\'e de Strasbourg/CNRS-IN2P3, 23 Rue du Loess, F-67037 Strasbourg, France\\
$^{(c)}$ Theory Division, Physics Department, CERN, CH-1211 Geneva 23, Switzerland}

\begin{abstract}
We investigate trilepton final states to probe top anomalous couplings at the Large Hadron
Collider. We focus on events originating from the associated production of a single top quark with
a $Z$-boson, a channel sensitive to several flavor-changing neutral interactions of top and
up/charm
quarks. In particular, we explore a way to access simultaneously their anomalous couplings to
$Z$-bosons and gluons and derive the discovery potential of trilepton final states to such
interactions with 20 fb$^{-1}$ of 8~TeV collisions. We show that effective coupling strengths of
${\cal O}
(0.1-1)$ TeV$^{-1}$ can be reached. Equivalently, branching fractions of top quarks into lighter quarks and gluons or Z-bosons can be constrained to be below
${\cal O}(0.1-1)$\%.
\end{abstract}

\pacs{12.38.Bx,12.60.-i,14.65.Ha}

\maketitle

\section{Introduction}

Since its discovery, the top quark, given its large mass, is generally considered as a
sensitive probe to new physics. In particular, its flavor-changing neutral couplings to gluons and
$Z$-bosons $gqt$ and $Zqt$ (with $q\!=\!u,c$) vanish at tree-level as consequences of the
unbroken QCD gauge symmetry and the Glashow-Iliopoulos-Maiani mechanism.
This also guarantees that in the Standard Model, these
interactions stay suppressed at higher orders.
Deviations from these predictions have therefore been widely searched for at hadron colliders in
flavor-changing neutral top decays \cite{Abe:1997fz,Aaltonen:2008ac,Abazov:2011qf,Aad:2012ij,%
Chatrchyan:2012hqa} and single top processes~\cite{Aaltonen:2008qr,Abazov:2007ev,%
Abazov:2010qk, Aad:2012gd}. Expressed in terms of branching ratios, current
constraints are given at the 95\% confidence level by $\mathcal{BR}(t\to ug) \!<\!
5.7 \cdot 10^{-3} \%$, $\mathcal{BR}(t\to cg)
\!<\! 2.1 \cdot 10^{-2}\%$ and $\mathcal{BR}(t\to qZ) \!<\! 0.21\%$,
assuming one single non-vanishing anomalous coupling.

Existing phenomenological analyses \cite{Hosch:1997gz,AguilarSaavedra:2004wm,Ferreira:2005dr,%
Liu:2005dp,Gao:2011fx,Li:2011ek,Gong:2013sh} mostly focus on direct top quark
production and, in a smaller extent, on rarer processes and indirect probes.
In this work, we concentrate on the associated production of a top
quark with a $Z$-boson. Although monotop signatures related to invisible $Z$-boson decays can be
interesting for new physics searches \cite{Andrea:2011ws}, we focus on final states
where both heavy particles decay into charged leptons. On the one hand, this
yields a trilepton topology inferring a Standard Model background under good
control. On the other hand, this allows us to
design an analysis probing at the same time $gqt$ and $Zqt$ interactions, in
contrast to standard investigations that are only sensitive to one type of couplings.
Since the production rate of a top quark in association with a $Z$-boson is
highly suppressed in the Standard Model, we aim to interpret possible excesses in trilepton events
as hints of large anomalous $gqt$ and $Zqt$ couplings, as predicted by many
new physics theories~\cite{delAguila:1999ec}.

In order to facilitate the analysis of new physics effects in top anomalous
couplings, the relevant interactions are usually described in terms of a minimal set of
effective operators independent of the underlying theory~\cite{AguilarSaavedra:2008zc}.
This bottom-up approach is in particular well motivated when new particles are heavy so that
they can be integrated out.
In this context, limits on new physics can be derived in a model-independent way. To this aim, we analyze, by means of Monte Carlo simulations, trilepton events to be produced at the Large Hadron Collider (LHC).

This work extends a pioneering study focusing on 14~TeV runs of the LHC with respective
luminosity of 10~fb$^{-1}$ and 100 fb$^{-1}$~\cite{AguilarSaavedra:2004wm}. First, we update the
older results according to current LHC settings, \ie, 20 fb$^{-1}$ of proton-proton
collisions with a center-of-mass energy of 8~TeV. Next, we make use of a more accurate description
of the Standard Model background relying on state-of-the-art Monte Carlo event
generation including multiparton matrix-element merging and on a more advanced simulation of the
detector response. We then
design a better adapted search strategy that we believe to be worthwhile to be tested
by the ATLAS and CMS experiments.

\section{Event simulation}
\begin{figure}[t]
  \SetScale{.27}\begin{picture}(808,57) (0,-5)
    \SetColor{Black}
    \Line[arrow,arrowpos=0.5,arrowlength=5,arrowwidth=2,arrowinset=0.2](144,69)(320,69)
    \Line[arrow,arrowpos=0.5,arrowlength=5,arrowwidth=2,arrowinset=0.2](32,165)(144,69)
    \Gluon(32,-27)(144,69){7.5}{12}
    \Line[arrow,arrowpos=0.5,arrowlength=5,arrowwidth=2,arrowinset=0.2](320,69)(432,165)
    \Photon(320,69)(432,-27){7.5}{7}
    \SetColor{OrangeRed}
    \Vertex(144,69){9.22}
    \Text(60,25)[lb]{\small{\Black{$t$}}}
    \Text(102,40)[lb]{\small{\Black{$t$}}}
    \Text(102,10)[lb]{\small{\Black{$Z$}}}
    \Text(15,40)[lb]{\small{\Black{$u/c$}}}
    \Text(15,5)[lb]{\small{\Black{$g$}}}
    \SetColor{Black}
    \Line[arrow,arrowpos=0.5,arrowlength=5,arrowwidth=2,arrowinset=0.2](584,69)(760,69)
    \Line[arrow,arrowpos=0.5,arrowlength=5,arrowwidth=2,arrowinset=0.2](472,165)(584,69)
    \Gluon(472,-27)(584,69){7.5}{12}
    \Line[arrow,arrowpos=0.5,arrowlength=5,arrowwidth=2,arrowinset=0.2](760,69)(872,165)
    \Photon(760,69)(872,-27){7.5}{7}
    \SetColor{OrangeRed}
    \Vertex(764,71){9.22}
    \SetColor{Black}
    \Text(182,25)[cb]{\small{\Black{$u/c$}}}
    \Text(222,40)[lb]{\small{\Black{$t$}}}
    \Text(222,10)[lb]{\small{\Black{$Z$}}}
    \Text(135,40)[lb]{\small{\Black{$u/c$}}}
    \Text(135,5)[lb]{\small{\Black{$g$}}}
 \end{picture}
  \caption{\label{fig:feynman} Representative Feynman diagrams leading to the production of a
    single top quark in association with a $Z$-boson. Flavor violation (red dots) occurs either in
    the strong (left) or weak (right) sector.}
\end{figure}
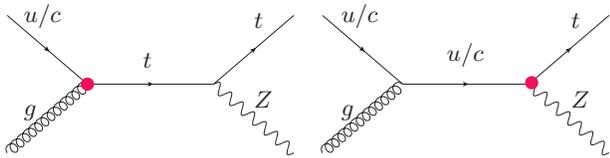

Single top production in association with a $Z$-boson can be driven, at tree-level, by two
main mechanisms related to the strong and weak sector, respectively. This is illustrated by the
two representative Feynman diagrams of Figure~\ref{fig:feynman} where flavor-changing interactions
of top and light\footnote{The naming light quark (jet) also refers to
charmed objects.} quark fields $t$ and $q$ involve either a gluon ($G^a_\mu$) or a $Z$-boson
($Z_\mu$). Both cases can be described by means of effective
operators to be supplemented to the Standard Model Lagrangian~\cite{Malkawi:1995dm,Hosch:1997gz,%
AguilarSaavedra:2008zc}. Denoting the fundamental
representation matrices of $SU(3)$ by $T_a$, the strong and weak coupling constants by $g_s$ and
$g$, the cosine of the weak mixing angle by $c_W$ and the $Z$-boson and gluon field strength
tensors by $Z_{\mu\nu}$ and $G^a_{\mu\nu}$, this effective Lagrangian reads
\be\bsp
  \mathcal{L} =&\ \sum_{q=u,c} \bigg[
    \sqrt{2} g_s \frac{\kappa_{gqt}}{\Lambda}\ \bar{t}\sigma^{\mu\nu}T_a
        (f^L_q P_L \!+\! f^R_q P_R) q\ G^a_{\mu\nu}\\
    &\ + \frac{g}{\sqrt{2} c_W} \frac{\kappa_{zqt}}{\Lambda}\  \bar{t}\sigma^{\mu\nu}
        (\hat f^L_q P_L \!+\! \hat f^R_q P_R) q\ Z_{\mu\nu} \\
    &\ + \frac{g}{4 c_W} \zeta_{zqt}\ \bar{t}\gamma^{\mu}
        (\tilde f^L_q P_L \!+\! \tilde f^R_q P_R) q\ Z_{\mu} \bigg]
   + \text{h.c.} \ .
\esp\label{eq:lag}\ee
In this expression, $P_L$ and $P_R$ are chirality projectors acting on spin space and
$\sigma^{\mu\nu} \!=\! \frac{i}{4}[\gamma^\mu, \gamma^\nu]$.
While other operators are in principle possible, they can always be reexpressed in terms of those
included in Eq.~\eqref{eq:lag} so that we only consider the minimal set of independent Lagrangian
terms above.
The magnitude of new physics effects that are assumed to appear at an energy scale $\Lambda$ are
modeled through real dimensionless parameters $\kappa_{gqt}$, $\kappa_{zqt}$ and $\zeta_{zqt}$
together with complex chiral
parameters $f^{L,R}_q$, $\hat f^{L, R}_q$ and $\tilde f^{L,R}_q$ normalized to $|f^L_q|^2 \!+\!
|f^R_q|^2 \!=\! |\hat f^L_q|^2 \!+\! |\hat f^R_q|^2 \!=\! |\tilde f^L_q|^2 \!+\! |\tilde f^R_q|^2
\!=\!1$.

In this work, we perform a phenomenological analysis based on Monte Carlo simulations of collisions produced at
the LHC, running at a center-of-mass energy of $\sqrt{s}=$8~TeV and for an integrated luminosity of 20~fb$^{-1}$.
For both signal and background,
we use the
\madgraph~5~\cite{Alwall:2011uj} package for the generation of hard scattering matrix elements
including up to two additional jets and convolved with the leading order set of
the CTEQ6 parton density fit \cite{Pumplin:2002vw}. Parton-level events produced in this way
are then matched to parton showering and hadronization by means of the program
\pythia~6~\cite{Sjostrand:2006za} and merged
according to the $k_T$-MLM scheme \cite{Mangano:2006rw,Alwall:2008qv}.
Detector effects are subsequently accounted for using a modified version
of \delphes~2.0~\cite{Ovyn:2009tx}. The latter includes a modeling of the performances of the CMS
detector as described in Ref.~\cite{Ball:2007zza} and a more recent description of the $b$-tagging
efficiency and mistagging rates. For the latter, we base our implementation on the {\sc  Tchel}
algorithm~\cite{CMS:2009gxa,CMS:2011cra}, which leads, \eg, to a $\sim\!75\%$ tagging
efficiency and a $10\!-\!15\%$ mistagging rate for jets with a transverse momentum
of $50\!-\!80$~GeV. After employing the {\sc FastJet}
package~\cite{Cacciari:2005hq} for jet reconstruction with an anti-$k_{t}$
algorithm of radius parameter set to $R=0.4$, simulated events are analyzed within
the \madanalysis~5~framework~\cite{Conte:2012fm}.

In order to allow for signal simulation, we have implemented the effective Lagrangian of
Eq.~\eqref{eq:lag} within the \feynrules~package~\cite{Christensen:2008py,Christensen:2009jx,%
Christensen:2010wz, Duhr:2011se,Fuks:2012im} and subsequently exported the model to a UFO
module~\cite{Degrande:2011ua} that has been linked to \madgraph~5.
We then generate events describing the production of a top (anti)quark decaying leptonically
(together with a $b$-jet and missing energy), in association with
a pair of same flavor leptons with opposite electric
charges\footnote{By the terminology \textit{leptons}, we equivalently denote electrons, muons and
taus decaying leptonically.} whose the invariant mass is greater than 12 GeV. In contrast to the diagrams of
Figure~\ref{fig:feynman}, virtual photon and off-shell $Z$-boson effects are included in the simulation.

In our study, we investigate
simplified scenarios with $f^L_{q} \!=\!\hat f^L_q\!=\! 
\tilde f_q^L \!=\! 0$ and $f^R_{q} \!=\! \hat f^R_q \!=\! \tilde f_q^R
\!=\! 1$. Leading order inclusive
cross sections $\sigma_{tZ}(x)$ read, assuming
a single non-vanishing anomalous coupling $x$ at a time,
\be\bsp
  \sigma_{tZ}(\kappa_{gut}/\Lambda) =&\ 86.78\ \big|\kappa_{gut} \
    (1 \text{TeV})/\Lambda\big|^2\ \text{pb} \ , \\
  \sigma_{tZ}(\kappa_{gct}/\Lambda) =&\ 3.255\ \big|\kappa_{gct}\
    (1 \text{TeV})/\Lambda\big|^2\ \text{pb} \ , \\
  \sigma_{tZ}(\kappa_{zut}/\Lambda) =&\ 5.769\ \big|\kappa_{zut}\
    (1 \text{TeV})/\Lambda\big|^2\ \text{pb} \ , \\
  \sigma_{tZ}(\kappa_{zct}/\Lambda) =&\ 0.273\ \big|\kappa_{zct}\
    (1 \text{TeV})/\Lambda\big|^2\ \text{pb} \ , \\
  \sigma_{tZ}(\zeta_{zut}) =&\ 2.727\cdot10^1\ \big| \zeta_{zut}\big|^2\ \text{pb}\ , \\
  \sigma_{tZ}(\zeta_{zct}) =&\ 1.533\phantom{\cdot10^1}\ \ 
    \big|\zeta_{zct}\big|^2\ \text{pb} \ .
\esp\label{eq:xsec}\ee
Those predictions agree with the results of Ref.~\cite{AguilarSaavedra:2010rx}.
In the following, we consider that each type of
interaction can be treated independently, \ie, that
either the strong ($\kappa_{gqt}$) or one of the
weak ($\kappa_{zqt}$ or $\zeta_{zqt}$) vertices are non-zero,
and we include next-to-leading order (NLO) $K$-factors fixed to 1.3 \cite{Gao:2011fx,Li:2011ek}.

We now turn to the simulation of the Standard Model background. First, we do not consider
multijet events since their correct treatment requires data-driven methods.
We instead rely on existing experimental analyses that have shown that they contribute negligibly
after a selection based on the trilepton topology and missing energy
requirements~\cite{Aad:2012twa,Chatrchyan:2012bd}.
Events originating from the production of a single gauge boson, decaying leptonically or invisibly,
together with jets have been normalized to the
next-to-next-to-leading order (NNLO), using total rates of 35678~pb and 10319~pb
for $W$- and $Z$-boson production
calculated by means of the {\sc Fewz} program~\cite{Gavin:2012sy,Gavin:2010az} with the CT10
parton densities~\cite{Lai:2010vv}. Inclusive top-antitop events have been normalized to
255.8~pb, as predicted by
the {\sc Hathor} package~\cite{Aliev:2010zk} after convolving all
NLO diagrams and genuine NNLO contributions with the CT10 parton
densities. Single top event generation has been split into the production of three
inclusive samples, normalized, at an approximate NNLO accuracy,
to 87.2~pb, 22.2~pb and 5.5~pb for the $t$, $tW$ and $s$ channels~\cite{Kidonakis:2012db}.
Diboson event weights have been rescaled according to NLO
predictions of 30.2~pb, 11.8~pb and 4.5~pb for the $WW$, $WZ$ and $ZZ$ modes, respectively, as
computed by means of the {\sc Mcfm} program~\cite{Campbell:1999ah,%
Campbell:2011bn,Campbell:2012dh}, employing again CT10 densities and after neglecting
full hadronic decay modes.
Finally, $ttV$ (with $V\!=\!W,Z$) events have been normalized to
NLO as predicted by {\sc Mcfm} while other simulated rare Standard
Model processes rely
on \madgraph~5 leading-order results. Inclusive cross sections for the
$ttW$, $ttZ$, $tZj$, $ttWW$ and $tttt$ channels have been found to be
0.25~pb, 0.21~pb, 46~fb, 13~fb and 0.7~fb, respectively.

\section{Anomalous top couplings at the LHC}
Events are preselected by requiring exactly three isolated charged leptons (electrons or muons)
with a transverse momentum $p_T \!\geq\! 20$~GeV and a pseudorapidity $|\eta|\!\leq\! 2.5$, using the CMS
coordinate system presented, \eg, in Ref.~\cite{Ball:2007zza}. Lepton isolation is enforced by optimizing,
independently for electrons and muons, an isolation variable $I_{\rm rel}$ so that the analysis sensitivity to
top anomalous couplings is maximized, the sensitivity being defined as the significance $S/\sqrt{S+B}$ where
$S$ and $B$ are respectively the number of signal and background events after all selections.
The quantity $I_{\rm rel}$ is derived from the amount of transverse energy, evaluated
relatively to the lepton $p_T$, present in a cone of radius $R \!=\! \sqrt{\Delta\varphi^2 +
\Delta\eta^2} \!=\! 0.3$ centered on the lepton, $\varphi$ being the azimuthal angle with
respect to the beam direction. The selected electrons and muons are asked to satisfy the constraint
$I_{\rm rel} < 0.28 $ and $I_{\rm rel}<0.06$, respectively.

Two leptons, labeled by $\ell_1$
and $\ell_2$, are tagged as originating from a $Z$-boson decay by selecting
the pair of same flavor leptons with opposite electric charges whose the invariant mass
$m_{\ell_1\ell_2}$ is the closest to the $Z$-boson mass $m_Z$. We further reject events for which
$|m_{\ell_1\ell_2} \!-\! m_Z| \! \geq \! 15$~GeV.

\begin{table}[!t]
\renewcommand{\arraystretch}{1.3}
\caption{\label{tab:cutflow}
  Number of expected events for 20 fb$^{-1}$ of LHC collisions at $\sqrt{s}=8$ TeV,
  together with the associated statistical uncertainties, after
  each of the selections described in the text. Two representative signal scenarios have been
  considered, with non-vanishing $\kappa_{gut}$ (second column) and $\kappa_{zut}$
  (third column) parameters, respectively, after enforcing a signal normalization of 10~fb. The
  sum over all background contributions is also indicated (last column).}
\begin{tabular}{|l|c|c|c|}
\hline
Selection                               & $\kappa_{gut} \neq 0$          & $\kappa_{zut} \neq 0$     &  Background  \\
\hline
Trilepton topology                      & 41.3$\pm$0.3 & 36.6$\pm$0.3& 3648.3$\pm$143.1 \\ 
$m_{\ell_1\ell_2} \!\in\! [76,106]$ GeV & 40.1$\pm$0.3 & 35.6$\pm$0.3& 3520.6$\pm$140.8 \\ 
$\met \geq 30$ GeV                      & 33.5$\pm$0.3 & 28.1$\pm$0.3& 1484.9$\pm$32.1  \\ 
$m_T^W \geq 10$ GeV                     & 31.1$\pm$0.2 & 26.6$\pm$0.2& 1373.9$\pm$19.2  \\ 
At least one jet                        & 26.5$\pm$0.2 & 23.4$\pm$0.2&  624.4$\pm$18.8  \\ 
One single $b$-tagged jet               & 18.3$\pm$0.2 & 15.4$\pm$0.2&  133.2$\pm$3.3   \\ 
$m^{(\text{reco})}_t \leq 250$ GeV      & 17.1$\pm$0.2 & 14.4$\pm$0.2&   75.3$\pm$2.7   \\ 
\hline
\end{tabular}
\renewcommand{\arraystretch}{1.0}
\end{table}

The third lepton $\ell_3$ is assumed to originate from a leptonically decaying top quark, being thus
accompanied by a neutrino yielding missing transverse energy \met~and by a jet issued from the
hadronization of a $b$-quark.
We therefore select events with $\met\!\geq\!30$~GeV and require the reconstructed $W$-boson transverse
mass  $m_T^W$ (as defined, \eg, in Ref.~\cite{Chatrchyan:2012hqa}) to fulfill $m_T^W \!\geq\!10$~GeV.
In addition, we focus on events containing at least one jet satisfying $p_T \!\geq\! 30$~GeV
and $|\eta| \!\leq\! 2.5$ and
exactly one $b$-tagged jet. For all jets, the ratio between
the hadronic and electromagnetic calorimeter deposits is also required to be larger than 30\%.
Finally, the four-momentum of the $W$-boson from which the missing energy and the lepton $\ell_3$ are
issued is reconstructed, using the $W$-boson mass as a constraint. The top quark
mass $m^{(\text{reco})}_t $ can then be determined by computing the invariant mass of the
$W$-boson and the $b$-tagged jet system, and further employed for background rejection. The
reconstructed top mass is asked to fulfill $m^{(\text{reco})}_t \!\leq\! 250$~GeV, this
criterion being found to maximize the analysis sensitivity to new physics.

The number of events surviving each of the selection criteria is indicated, together with the
associated statistical uncertainties, in
Table~\ref{tab:cutflow} for both the Standard Model background and two representative signal
scenarios for which the only non-vanishing anomalous coupling is respectively $\kappa_{gut}$
and $\kappa_{zut}$. After all selections, $36.4\pm0.7$ diboson, $37.1\pm2.6$ $t\bar{t}$
and $1.3\pm0.1$ $tZj$ events remain, all other background contributions being negligible. In comparison, $17.1\pm0.2$ and
$14.4 \pm 0.2$ events are expected for the two signal scenarios after
normalizing both samples to a cross section of 10~fb, the difference being due to a different acceptance.

\begin{figure}[!t]
\centering
  \hspace{-.6cm}\includegraphics[width=.86\columnwidth]{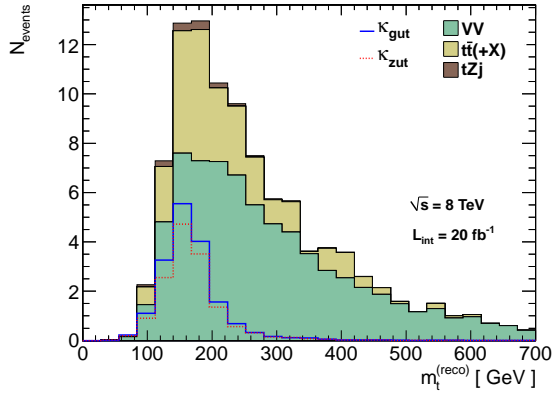}
  \caption{\label{fig:mtop}
    Distribution of the reconstructed top mass $m_t^{(\text{reco})}$ after the selection strategy
    described in the text, with the exception of the final cut, for 20
    fb$^{-1}$ of LHC collisions at $\sqrt{s}=$8~TeV. We indicate the dominant diboson ($VV$), top-antitop
    pair with possibly one or several gauge bosons ($t \bar t (+X)$) and $tZj$
    contributions to the Standard Model background, as well as two signal spectra
    for two representative new physics scenarios, normalized to 10~fb.}
\end{figure}

The effect of the top reconstruction is illustrated on Figure \ref{fig:mtop}. After applying the
selection strategy described above but the last step, we first present the $m_t^{(\text{reco})}$
spectrum in the Standard Model.
The predictions show a broad peaky distribution, with a maximum close to the top mass.
We then superimpose two representative signal distributions, normalized to 10~fb.
They all present narrower peaks, the distribution maximum being at the top mass.
Confronting the shapes of the two signal spectra also indicates
that observing $tZ$ trilepton events does not provide any
discriminating power among weak and strong flavor-changing neutral
top interactions.

\begin{figure}[!t]
\centering
  \hspace{-.6cm}\includegraphics[width=.86\columnwidth]{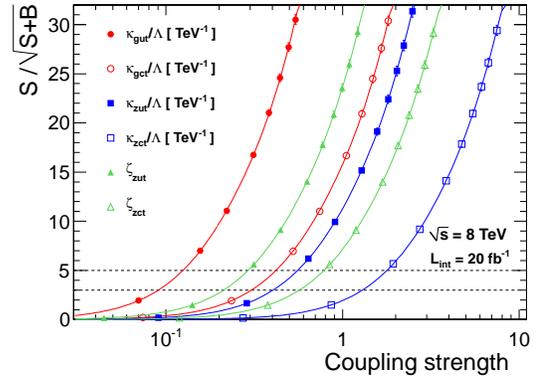}
  \caption{\label{fig:sign1D}
    LHC sensitivity to the considered top anomalous couplings as a function of the
    interaction strengths, given as $S/\sqrt{S+B}$ after applying the analysis
    described in the text.}
\end{figure}

The sensitivity of the 2012 LHC run to the four top anomalous couplings under consideration,
defined as $S/\sqrt{S+B}$ (see above), is presented for several choices of the
anomalous coupling
parameters in Figure~\ref{fig:sign1D} in the case we assume a single non-vanishing coupling at a
time. The results are subsequently fitted by polynomial functions so that
$3\sigma$ and $5\sigma$ discovery ranges are extracted. Hence, anomalous couplings such as
\be\bsp
& \kappa_{gut}/\Lambda \geq 0.12 \text{ TeV}^{-1} \ (0.09 \text{ TeV}^{-1})\ , \\
& \kappa_{gct}/\Lambda \geq 0.42 \text{ TeV}^{-1} \ (0.31 \text{ TeV}^{-1})\ , \\
& \kappa_{zut}/\Lambda \geq 0.56 \text{ TeV}^{-1} \ (0.40 \text{ TeV}^{-1})\ , \\
& \kappa_{zct}/\Lambda \geq 1.78 \text{ TeV}^{-1} \ (1.30 \text{ TeV}^{-1})\ , \\
& \zeta_{zut} \geq 0.29  \ (0.21)\ , \\
& \zeta_{zct} \geq 0.78 \ (0.57)\ ,
\esp \ee
can be reached at the $5\sigma$ ($3\sigma$) level with 20 fb$^{-1}$ of 8~TeV LHC collisions.
Using NNLO results for the decay width of the top quark in the Standard Model
\cite{Gao:2012ja}, the limits can be translated in terms of constraints on the branching
fractions of rare top decays,
\be\bsp
  & \mathcal{B}\mathcal{R}(t \to gu) \leq 0.47\%\ (0.25 \%)\ , \\
  & \mathcal{B}\mathcal{R}(t \to gc) \leq \phantom{0}5.1\%\ (2.8 \%)\ , \\
  & \mathcal{B}\mathcal{R}(t \to Zu) \leq 0.39\%\ (0.20\%)           \quad\text{with}\ \kappa_{zut}/\Lambda\neq 0\ , \\
  & \mathcal{B}\mathcal{R}(t \to Zc) \leq \phantom{0}3.8\%\ (2.1 \%) \quad\text{with}\ \kappa_{zct}/\Lambda\neq 0\ , \\
  & \mathcal{B}\mathcal{R}(t \to Zu) \leq 1.07\%\ (0.56\%)           \quad\text{with}\ \zeta_{zut}\neq 0 \ , \\
  & \mathcal{B}\mathcal{R}(t \to Zc) \leq \phantom{0}7.2\%\ (4.0 \%) \quad\text{with}\ \zeta_{zct}\neq 0 \ ,
\esp\label{eq:BR}\ee
at the $5\sigma$ ($3\sigma$) levels, the non-vanishing anomalous coupling being indicated
for the weak decay cases.
The obtained limits on $\mathcal{B}\mathcal{R}(t \!\to\! gq / Zc)$
are found to be not competitive by one or two orders of magnitude with results extracted from
direct top production and flavor-changing top decays~\cite{Aad:2012gd,Chatrchyan:2012hqa}.
In contrast, the bounds obtained on $\mathcal{B}\mathcal{R}(t \!\to\! Zu)$ are promising. First, they show
that the trilepton analysis which we are proposing can be used to confirm (or possibly improve) the
current constraints extracted from flavor-changing top decays in $t \bar t$ events at $\sqrt{s} =
7$~TeV \cite{Chatrchyan:2012hqa,Aad:2012ij} by means of an independent measurement at a different
center-of-mass energy an with a different integrated luminosity. Next, both channels are statistically
independent and thus valuable to be combined to get stronger bounds on the $Ztu$ coupling strength.

\begin{figure}[!t]
  \centering
  \hspace{-.6cm}\includegraphics[width=.86\columnwidth]{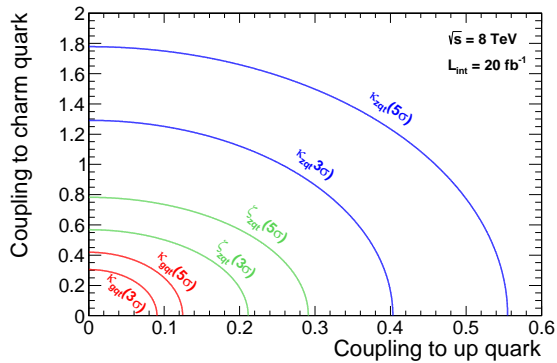}
  \caption{\label{fig:sign2D}
    $3\sigma$ and
    $5\sigma$ discovery ranges for top anomalous interactions in the strong
    (blue, non-zero $\kappa_{gut}/\Lambda$ and $\kappa_{gct}/\Lambda$) and
    weak (red, non-zero $\kappa_{zut}/\Lambda$ and $\kappa_{zct}/\Lambda$;
    green, non-zero $\zeta_{zut}$ and $\zeta_{zct}$)
    sector.}
\end{figure}

We now allow for several non-zero couplings simultaneously, either in the strong sector
(non-vanishing $\kappa_{gut}/\Lambda$ and $\kappa_{gct}/\Lambda$ parameters) or in the weak
sector (non-vanishing $\kappa_{zut}/\Lambda$ and $\kappa_{zct}/\Lambda$ or
non-vanishing $\zeta_{zut}$ and $\zeta_{zct}$ parameters). In
Figure~\ref{fig:sign2D}, we extract the associated $3\sigma$ and $5\sigma$ discovery reaches. We
observe a better sensitivity to flavor-changing
interactions with an up quark than with a charm quark, as expected from
parton densities, the charm content of the proton being suppressed with respect to its
up content.

\section{Conclusions}
In this letter, we have investigated the discovery potential of trilepton final states
to anomalous, flavor-changing, neutral couplings of the top and light quarks to
gluon and $Z$-boson fields. These interactions induce possibly significant production rates of such
final states via the associated production
of a leptonically decaying top quark with a dilepton pair.
We have designed a search strategy capable to probe simultaneously $gqt$ and $Zqt$
couplings and showed that the 2012 LHC run can reach interaction strengths of order
$0.1-1$~TeV$^{-1}$. Equivalently, bounds on the associated rare top branching
fractions as low as ${\cal O}(0.1-1)\%$ could be set, possibly improving, in particular, the
current limits on the
$t\!\to\!Zu$ decay. Our results therefore motivate a further investigation by the ATLAS
and CMS collaborations in the context of real LHC data.

\begin{acknowledgements}
The authors are grateful to C.~Collard for discussions and a careful reading of the manuscript. This
work has been partially supported by the Theory-LHC France-initiative of the CNRS/IN2P3 and
by the French ANR 12 JS05 002 01 BATS@LHC.
\end{acknowledgements}

\bibliography{review}

\end{document}